\documentclass[journal,monochrome]{IEEEtran}

\def\baselinestretch{0.95}

\usepackage{cite}
\ifCLASSINFOpdf
\usepackage[pdftex]{graphicx}
\DeclareGraphicsExtensions{.pdf,.jpeg,.png}
\else
\usepackage[dvips]{graphicx}
\DeclareGraphicsExtensions{.eps}
\fi
\usepackage{epsfig,epsf,epstopdf,graphicx}
\usepackage[cmex10]{amsmath}
\usepackage{amssymb}
\usepackage{amsthm }
\usepackage{nicefrac}
\usepackage{hyperref}
\usepackage{xcolor}
\hypersetup{colorlinks=true}
\usepackage{tabularx}

\theoremstyle{theorem}

\theoremstyle{definition}

\theoremstyle{plain}

\theoremstyle{plain}

\newcommand\Tstrut{\rule{0pt}{2.6ex}}         % = `top' strut
\newcommand{\Bb}{{\textbf{B}}}

\newcommand{\Gb}{{\textbf{G}}}
\newcommand{\ub}{{\textbf{u}}}

\newcommand{\fb}{{\textbf{f}}}

\newcommand{\gb}{{\textbf{g}}}

\newcommand{\Ab}{{\textbf{A}}}
\newcommand{\nb}{{\textbf{n}}}

\newcommand{\Ib}{{\textbf{I}}}

\newcommand{\rb}{{\textbf{r}}}
\newcommand{\Rb}{{\textbf{R}}}
\newcommand{\hb}{{\textbf{h}}}

\newcommand{\Vb}{{\textbf{V}}}

\newcommand{\pb}{{\textbf{p}}}

\newcommand{\phib}{{\mbox{\boldmath $\phi$}}}

\newcommand{\Deltab}{{\mbox{\boldmath $\Delta$}}}
\newcommand{\etab}{{\mbox{\boldmath $\eta$}}}

\newcommand{\omegab}{{\mbox{\boldmath $\omega$}}}

\newcommand{\nub}{{\mbox{\boldmath $\nu$}}}

\usepackage{algorithm}
\usepackage{multirow}
\usepackage{algcompatible}
\usepackage[bottom]{footmisc}
\usepackage{tablefootnote}
\usepackage{tabularx}

\PassOptionsToPackage{dvipsnames}{xcolor}
\usepackage{tikz}
\usetikzlibrary{calc}

\ifCLASSOPTIONcompsoc
  \usepackage[caption=false,font=normalsize,labelfont=sf,textfont=sf]{subfig}
\else
  \usepackage[caption=false,font=footnotesize]{subfig}
\fi
\usepackage{fixltx2e}
\usepackage{afterpage}
\usepackage{float}
\usepackage{stfloats}

\makeatletter

\begin{document}

\title{Double-Private Distributed Estimation Algorithm Using Differential Privacy and a Key-Like Proportionate Matrix with Its Performance Analysis}

\author{Mehdi Korki, \IEEEmembership{Member,~IEEE,}
Fatemehsadat Hosseiniamin, \IEEEmembership{Student Member,~IEEE,}
Hadi~Zayyani, \IEEEmembership{Member,~IEEE,} and Mehdi Bekrani
%~\IEEEmembership{Member,~IEEE,}

\thanks{M.~Korki is with the School
of Science, Computing, and Engineering Technologies, Swinburne University of Technology, Melbourne, Australia (e-mail: mkorki@swin.edu.au).}% <-this %
\thanks{F. Hosseiniamin, H.~Zayyani, and M. Bekrani are with the Department
of Electrical and Computer Engineering, Qom University of Technology (QUT), Qom, Iran (e-mails: zayyani@qut.ac.ir, hosseiniamin110@gmail.com, bekrani@qut.ac.ir).}% <-this %
\vspace{-0.5cm}}


\maketitle
\thispagestyle{plain}
\pagestyle{plain}

\begin{abstract}
In this brief, we present an enhanced privacy-preserving distributed estimation algorithm, referred to as the ``Double-Private Algorithm," which combines the principles of both differential privacy (DP) and cryptography. The proposed algorithm enhances privacy by introducing DP noise into the intermediate estimations of neighboring nodes. Additionally, we employ an inverse of a closed-form reproducible proportionate gain matrix as the cryptographic key matrix to fortify the privacy protection within the proposed double private algorithm. \textcolor{blue}{We improve the algorithm by transmitting alternative variable vectors instead of raw measurements, resulting in enhanced key matrix reconstruction performance. This innovative approach mitigate noise impact, enhancing overall algorithm effectiveness.} We also establish an upper bound for the norm of the error between the non-private Diffusion Least Mean Square (DLMS) algorithm and our double private algorithm. Further, we determine a sufficient condition for the step-size to ensure the mean convergence of the proposed algorithm. Simulation results demonstrate the effectiveness of the proposed algorithm, particularly its ability to attain the final Mean Square Deviation (MSD) comparable to that of the non-private DLMS.

\end{abstract}


% \fi
\IEEEpeerreviewmaketitle

\section{Introduction}
\label{sec:Intro}
\IEEEPARstart{T}{he} distributed estimation finds applications in a diverse range of multi agent network scenarios such as Wireless Sensor Networks (WSN), communication networks, and biological networks \cite{Sayed14}. The multiple agents or nodes of the network can collaborate to generate estimations of an unknown vector. Collaboration can be achieved through various strategies, including incremental, consensus, and diffusion methods. Among these, the diffusion approach stands out for its versatility, scalability, minimal storage requirements, and ease of implementation \cite{Sayed14}, \cite{Tu12}.

In the context of the Adapt-Then-Combine (ATC) diffusion algorithms \cite{Kumar21}-\cite{Amini23}, the process unfolds in two distinct steps. Initially, agents update their individual estimates based on prior estimations and current measurements, guided by a local cost function —an operation referred to as the adaptation step. During this phase, each local agent accesses its own data while preserving node-level privacy.
Subsequently, in the combination step, neighboring agents collaborate with the local agent by sharing their own estimations. This collaborative effort is aimed at refining the overall estimate. However, this step introduces privacy concerns. In the presence of malicious or adversarial agents, there is a potential for unauthorized access to the global estimation of the network. Consequently, a robust privacy-preserving strategy is essential to protect against eavesdropping. In essence, a solution is required to ensure secure transmission between network agents.

%The secure solution may be cryptographic method which needs to exchange the keys. This increases the communication load of the network and results in a demanding power for communications and even computations \cite{Lage13}. But, if there is a mechanism for reaching a common key without communication is a good solution for this category and prohibit of eavesdropping of the key by the adversaries. Another solution is simple methods like noise-injecting algorithms \cite{He19}-\cite{SadeK22}. Among these methods, the differential privacy technique \cite{He20}-\cite{SadeK22} is the common method which inject the uncorrelated noise to the shared information signal to guarantee the privacy. Various noises are used in this regard which Gaussian, Laplacian, and offset-symmetric Gaussian are some examples \cite{SadeK22}. Moreover, a few privacy-preserving diffusion algorithms are suggested in the literature \cite{Gogi22}-\cite{Rizk23}. In \cite{Gogi22}, a private-partial distributed DLMS is proposed in which the combination step is replaced by a average consensus using a perturbed noise. Recently, \cite{Rizk23} suggests differential privacy schemes in which the noise is added in all steps of diffusion algorithm.

The secure solution may involve a cryptographic method that requires key exchange. This key exchange process can significantly increase the communication load over the network, demanding considerable power for both communication and computations, as noted in \cite{Lage13}. An alternative approach that does not require constant communication for key exchange can be highly effective in preventing eavesdropping by adversaries. Another viable solution includes simple methods, such as noise-injecting algorithms mentioned in \cite{He19}-\cite{SadeK22}. Among these methods, differential privacy (DP) techniques~\cite{He20}-\cite{SadeK22} are widely employed. These techniques involve injecting uncorrelated noise into the shared information signal to ensure privacy. Various noise types, including Gaussian, Laplacian, and offset-symmetric Gaussian (OSGT), are commonly used in this context, as discussed in \cite{SadeK22}.

Furthermore, the literature suggests several privacy-preserving diffusion algorithms~\cite{Gogi22}-\cite{Wang21}. In \cite{Gogi22}, a private-partial distributed LMS is proposed, where the combination step is replaced by an average consensus method using perturbed noise. More recently,  the authors in \cite{Rizk23} introduced DP schemes that incorporate noise at all stages of the diffusion algorithm. Further, in \cite{Wang21}, the authors develop a privacy-preserving distributed projection least mean squares (LMS) strategy for linear multitask networks, where agents aim to enhance their local inference performance while protecting individual task privacy. It involves sending noisy estimates to neighbors, with the noise level optimized to balance accuracy and privacy.

%\cite{BkZy}

%In this brief, to ensure a higher level of privacy, we use both above mentioned techniques. So, we nominate it as a double private algorithm. At first, it uses a differential privacy technique and adds a noise to the intermediate estimations. At the second step, it uses a cryptography-like method and uses a Key matrix to multiply to the intermediate estimations. In this case, even if the enemy can have access to the differential-privacy noise, it needs to know the key matrix. On the other hand, if the adversary can reach to the key, they also need to know the noise sequence.

In this brief, we aim to enhance privacy through the simultaneous application of both aforementioned techniques, which is called a double private algorithm. Initially, we employ a differential privacy technique by introducing noise into the intermediate estimations. Subsequently, in the second phase, we leverage a cryptographic-like approach, utilizing a key matrix to perform multiplication on the intermediate estimations. This approach offers a dual layer of security: even if an adversary gain access to the differential-privacy noise, they would still require the knowledge of the key matrix. Conversely, if the adversary manages to obtain the key, they would additionally need to decipher the noise sequence. \textcolor{blue}{The proposed algorithm in \cite{Rizk23} emphasizes both guaranteed performance and privacy in distributed learning, while our double private algorithm enhances privacy with simultaneous privacy-preserving mechanisms, albeit without explicit performance guarantees.}

%The paper makes another contribution in preventing the distribution of the key matrix from the transmitting node (neighborhood node) to the receiving node (local node). This is achieved by employing the proportionate generalized correntropy-based diffusion LMS algorithm recently introduced in \cite{Amini23}. The significant advantage of using the proportionate gain matrix is its closed-form nature, which enables the local node to independently recreate the key matrix. Consequently, this approach eliminates the necessity of transmitting the key, reducing communication overhead and safeguarding key privacy from potential adversaries.

Another noteworthy advantage of the proposed double private scheme lies in its simplicity when it comes to encryption and decryption. This simplicity arises from the fact that both the proportionate gain matrix and its inverse are diagonal matrices. \textcolor{blue}{We have also implemented a novel approach to mitigate the impact of noise on the reconstruction of the key matrix. Specifically, we propose sending an alternative variable vector instead of raw measurements and regression vectors. Through this innovative modification, we have observed a notable enhancement in the performance of matrix reconstruction, consequently leading to improved overall performance of the proposed algorithm.}

Furthermore, the paper provides two essential mathematical analyses of the proposed method. The first analysis calculates an upper bound for the $l_2$-norm of the error between the non-private estimation and double private estimation. The second analysis establishes a sufficient condition for the step-size value to ensure the mean convergence of the proposed algorithm. Simulation results demonstrate that the proposed double-private algorithm can attain the final mean square deviation (MSD) comparable to that of the non-private DLMS algorithm with a delay.

\section{System Model and Problem Formulation}
\label{sec:ProblemForm}
A network topology of $N$ agents (nodes) is assumed in which the $k$'th agent collaborate with its neighborhood nodes collected in $\mathcal N_k$ encompassing itself. The $k$'th agent observes a linear measurement $d_{k,i}$ of an unknown $L\times 1$ vector denoted by $\omegab^o$ as $d_{k,i}=\ub_{k,i}^T\omegab^o+v_{k,i}$, where $i$ is the discrete time index, $\ub_{k,i}$ is the known $L\times 1$ regression vector, and $v_{k,i}$ is the measurement noise of $k$'th agent at time index $i$.

%\begin{equation}
%d_{k,i}=\ub_{k,i}^T\omegab^o+v_{k,i},
%\end{equation}
In a privacy-preserving ATC diffusion algorithm, there are two steps of adaptation and combination. In the adaptation step, the intermediate estimations are computed as $\phib_{k,i}=\omegab_{k,i}+\mu_l\pb_{k,i}$, where $\omegab_{k,i}$ is the estimation of node $k$ at the end of index $i$, and $\pb_{k,i}\triangleq\sum_{l\in \mathcal N_k}c_{l,k}\ub^T_{l,i}\mathrm{f}(d_{l,i}-\ub^T_{l,i}\omegab_{k,i})$ where function $\mathrm{f}(.)$ is dependent on the local cost function defined for the algorithm. For example, in classical DLMS algorithm, this function is $\mathrm{f}(x)=x$. In an uncooperative scenario, some adversary agents try to inject false data to abrupt the process of distributed estimation or at least eavesdrop the intermediate estimations and reach to the final estimate of unknown vector. By privacy preserving distributed estimation algorithms, we want to prevent them to access to the true estimations. So, for preserving the privacy, a disturbed (or encrypted) version of $\phib_{k,i}$ which is nominated by $\tilde{\phib}_{k,i}$ is transmitted to the neighbors by assuming AWGN channels between nodes. At the combination step, the perturbed intermediate estimation $\tilde{\phib}_{k,i}$ plus noise is received by the neighbors. Then, after de-perturbing (or decryption), the de-perturbed version of intermediate estimation which is $\tilde{\tilde{\phib}}_{l,i}$ is received by agent $k$ which is combined as $\omegab_{k,i+1}=\sum_{l\in \mathcal N_k}a_{l,k}\tilde{\tilde{\phib}}_{l,i}$.
%\begin{equation}
%\omegab_{k,i+1}=\sum_{l\in \mathcal N_k}a_{l,k}\tilde{\tilde{\phib}}_{l,i}.
%\end{equation}
The aim of privacy-preserving diffusion algorithm is to devise a well distributed algorithm with high privacy as possible.

\section{The Proposed double private proportionate generalized correntropy-based Diffusion LMS Algorithm}
\label{sec: prop}
In this section, we first explain the proportionate generalized correntropy-based diffusion LMS algorithm (PGCDLMS) since the proposed double-private algorithm is an extension of this algorithm. Then, the proposed double-private PGCDLMS (DP-PGCDLMS) is explained.

\subsection{Proportionate generalized correntropy-based Diffusion LMS Algorithm}
The PGCDLMS is essentially a proportionate DLMS (PDLMS) algorithm in which the proportionate gain matrix is obtained in a closed form \cite{Amini23}. So, the adaptation of PGCDLMS is as follows:
\begin{equation}
\phib_{k,i}=\omegab_{k,i}+\mu_k\Gb_k\pb_{k,i},
\end{equation}
where $\pb_{k,i}=[p_{k,i,1},...,p_{k,i,L}]^T=\sum_{l\in \cal N_\mathrm{k}}c_{lk}\ub_{l,i}(d_l(i)-\ub^T_{l,i}\omegab_{k,i})$ and the gain matrix $\Gb_k$ is obtained by optimizing a cost function defined by generalized correntropy \cite{Amini23}. The advantage of PGCDLMS algorithm is that there is an optimum closed-form formula for the gain matrix $\Gb_{k,i}=\mathrm{diag}(g^*_{k,i,r})$ which is
\begin{align}
\label{eq: g*}
g^{*}_{k,i,r}=\Big{[}\frac{\beta}{\mu_{k}}\Big{(}-\mathrm{Ln}(\frac{\lambda_{k,i}\beta^\alpha}{\mu_{k}^\alpha A}|p_{k,i,r}|^{-\alpha})\Big{)}^{1/\alpha}p_{k,i,r}^{-1}\Big{]},
\end{align}
where  $\alpha$ is the exponent parameter of generalized correntropy kernel of the algorithm, $\beta$ is the bandwidth parameter of the correntropy, and $\lambda_{k,i}=\lambda^{*}_{k,i}$ is
\begin{equation}
\label{eq.lambda}
\lambda^{*}_{k,i}=\exp\Big{(}\frac{1+\sum_{r}\big{(}\frac{\beta^\alpha}{\mu_{k}^\alpha}\mathrm{Ln}(\frac{\beta^\alpha}{\mu_{k}^\alpha A}|p_{k,i,r}|^{-\alpha})p_{k,i,r}^{-\alpha})}{\sum_{r}(-\frac{\beta^\alpha}{\mu_{k}^\alpha}p_{k,i,r}^{-\alpha})}\Big{)}.
\end{equation}

\subsection{The proposed double-private version of PGCDLMS}
The basic idea of DP-PGCDLMS is to use $\Gb^{-1}_{k,i}$ as the perturbation matrix which is multiplied by the intermediate estimation vector and then adding the differential privacy noise to that. So, we have
\begin{equation}
\label{eq: encvec}
\tilde{\phib}_{k,i}=\Gb^{-1}_{k,i}\phib_{k,i}+\etab_{k,i},
\end{equation}
where $\tilde{\phib}_{k,i}$ is the double private intermediate vector, $\etab_{k,i}$ is the differential privacy noise, and $\Gb^{-1}_{k,i}$ is the inverse of proportionate diagonal gain matrix used as a key matrix to perturb the data of intermediate estimation. In fact, the perturbation is an encryption mechanism to hide the intermediate estimation from unauthorized adversary agent which want to have access to the estimation. The perturbed vector $\tilde{\phib}_{l,i}$ of neighbor nodes is transmitted to the local node of $k$. So, the received perturbed vector of node $k$ from node $l$, assuming an AWGN channel between nodes, is $\rb_{l,i}=\tilde{\phib}_{l,i}+\hb_{l,i}$, where $\hb_{l,i}$ is the received noise vector. Then, the decrypted intermediate estimation is defined as
\begin{equation}
\label{eq: decvec}
\tilde{\tilde{\phib}}_{l,i}=\tilde{\Gb}_{l,i}(\rb_{l,i}-\etab_{l,i}),
\end{equation}
where $\tilde{\Gb}_{l,i}$ is the reconstructed key matrix, and it is assumed that the differential privacy noise $\etab_{l,i}$ is known for all honest agents. The noise generators are often implemented by an Linear feedback shift registers (LFSR) which is started by an initial condition. It is not difficult to set all the honest agents have the same LFSR and same initial conditions. So, the point is that adding a noise which is not known for adversaries, increase the privacy of the algorithm. It is one aspect of privacy preserving mechanism in our proposed double private scheme. The other aspect is to use a Key-like gain matrix to perturb the intermediate estimation. If the privacy noise is eavesdropped, then we have another second privacy preserving mechanism. There are \textcolor{blue}{three} cases for the reconstructed key matrix $\tilde{\Gb}_{l,i}=\hat{\Gb}_{l,i}+\Vb_{l,i}$, where $\Vb_{l,i}$ is the key error matrix. In first case, the reconstructed key matrix is approximately the true gain matrix i.e. $\tilde{\Gb}_{l,i}=\Gb_{l,i}+\Vb_{l,i}$ which can be obtained by, for example sharing the key matrix beforehand ($\Vb_{l,i}=0$) or transmitting the key matrix between nodes. As it is expected, this case is not practical because of high communication load for transmitting the key matrix or because of the danger of eavesdropping by adversaries. So, this case which we nominate the corresponding algorithm as oracle-DP-PGCDLMS is not practically feasible. But, we use it as a reference of comparisons. In the second case, the reconstructed key matrix $\tilde{\Gb}_{l,i}=\hat{\Gb}_{l,i}=\mathrm{diag}(\hat{g}_{l,i,r})$ is obtained by (\ref{eq: g*}) using  $\hat{\pb}_{l,i}=\sum_{l^{'}\in \cal N_\mathrm{k}}c_{l^{'},k}\ub_{l^{'},i}(d_{l^{'}}(i)-\ub^T_{l^{'},i}\omegab_{l,i})$. \textcolor{blue}{In this scenario, the vector $\pb_{l,i}$ is reconstructed using the noisy exchanges $d_{l'}(i)$ and $\ub_{l',i}$. Consequently, the noisy nature of these variables results in a noisy version of $\hat{\pb}_{l,i}$, thereby further amplifying the noise in the reconstructed key matrix. We designate this approach as DP-PGCDLMS-Version1. Alternatively, in the third case, we propose a direct exchange of $\pb_{l,i}$ at the $k$-th node. The noisy version of $\hat{\pb}_{l,i}=\pb_{l,i}+\nub_{l,i}$ solely impacts the reconstruction of the key matrix, leading to a denoised version of the proposed algorithm. We identify this modified version as DP-PGCDLMS-Version2.}
The order of computational complexity plus extra communication loads of the proposed DP-PGCDLMS \textcolor{blue}{algorithms (version 1 and version 2)} in comparison to some others are shown in Table 1. It is seen that the proposed algorithms are slightly more complex (the order of complexity is the same) than other algorithms and they need more communication load. \textcolor{blue}{Also, the DP-PGCDLMS-Version2 needs more communication loads in comparison to DP-PGCDLMS-Version1.}

\begin{table}[!b]
\caption{Computational Complexity per node $k$ and per iteration of algorithms ($N_k=\mathrm{Card}\{\cal N_\mathrm{k}\}$)}
\label{tab1}
\centering
 \begin{tabular}{p{19mm}||p{9mm}|p{20mm}|p{10mm}|p{16mm}} \hline
 \Tstrut Algorithm & Add  & Multiplication + Computation of G  & Extra Comm. load  \\\hline \hline

 \Tstrut DLMS \cite{LopS08}
 & $\!\!\! \begin{aligned} O(LN_k)& \end{aligned} $	
 & $\!\!\!\begin{aligned} Q(LN_k)& \end{aligned}$	
 %& $\!\!\!\begin{aligned} 0& \end{aligned}$	
 & $\!\!\!\begin{aligned} 0& \end{aligned}$ \\ \hline	
 %&	$\!\!\!\begin{aligned} N& \end{aligned}$  \\ \hline

 \Tstrut PR-DLMS \cite{Yim15}
 &	$\!\!\!\begin{aligned} O(LN_k)& \end{aligned}$	
 & $\!\!\!\begin{aligned} O(LN_k)+O(L^2)& \end{aligned}$
 %& $\!\!\!\begin{aligned} 0& \end{aligned}$
 & $\!\!\!\begin{aligned} 0& \end{aligned}$\\ \hline
 %& $\!\!\!\begin{aligned} N& \end{aligned}$ \\ \hline

%\Tstrut RG-PR-DLMS \cite{ZayyJ21}
% &	$\!\!\!\begin{aligned} L(3N_k-1)& \end{aligned}$	
% & $\!\!\!\begin{aligned} L(4N_k+2)& \end{aligned}$
% & $\!\!\!\begin{aligned} 0& \end{aligned}$
% & $\!\!\!\begin{aligned} O(L^3)\tablefootnote{Due to matrix inversion}& \end{aligned}$ \cite{ZayyJ21}\\ \hline
% %& $\!\!\!\begin{aligned} N& \end{aligned}$ \\ \hline

%\Tstrut DMCC \cite{Ma16}
% &	$\!\!\!\begin{aligned} L(3N_k-1)& \end{aligned}$	
% & $\!\!\!\begin{aligned} (3L+3)N_k& \end{aligned}$
% & $\!\!\!\begin{aligned} N_k& \end{aligned}$
% & $\!\!\!\begin{aligned} 0& \end{aligned}$ \\ \hline
% %& $\!\!\!\begin{aligned} N& \end{aligned}$ \\ \hline

%\Tstrut DFCAF \cite{Gogi20}
% &	$\!\!\!\begin{aligned} L(N_k+3)& \end{aligned}$	
% & $\!\!\!\begin{aligned} L(N_k+3)+\\2& \end{aligned}$
% & $\!\!\!\begin{aligned} 3& \end{aligned}$
% & $\!\!\!\begin{aligned} 0& \end{aligned}$ \\ \hline
% %& $\!\!\!\begin{aligned} N& \end{aligned}$ \\ \hline

\Tstrut PGCDLMS \cite{Amini23}
 &	$\!\!\!\begin{aligned} O(LN_k)& \end{aligned}$	
 & $\!\!\!\begin{aligned} O(LN_k)+O(L)& \end{aligned}$
 %& $\!\!\!\begin{aligned} 0& \end{aligned}$
 & $\!\!\!\begin{aligned} 0& \end{aligned}$ \\ \hline
 %& $\!\!\!\begin{aligned} N& \end{aligned}$ \\ \hline

\Tstrut \textcolor{blue}{DP-PGCDLMS-Version1}
 &	$\!\!\!\begin{aligned} O(LN_k)\\+O(L)& \end{aligned}$	
 & $\!\!\!\begin{aligned} O(LN_k)+O(2L)& \end{aligned}$
 %& $\!\!\!\begin{aligned} 0& \end{aligned}$
 & $\!\!\!\begin{aligned} O(LN_k) \end{aligned}$ \\ \hline
 %& $\!\!\!\begin{aligned} N& \end{aligned}$ \\ \hline

\Tstrut \textcolor{blue}{DP-PGCDLMS-Version2}
 &	$\!\!\!\begin{aligned} O(LN_k)\\+O(L)& \end{aligned}$	
 & $\!\!\!\begin{aligned} O(LN_k)+O(2L)& \end{aligned}$
 %& $\!\!\!\begin{aligned} 0& \end{aligned}$
 & $\!\!\!\begin{aligned} O(2LN_k) \end{aligned}$ \\ \hline
 %& $\!\!\!\begin{aligned} N& \end{aligned}$ \\ \hline

\end{tabular}

\end{table}

\section{Mathematical analysis}
\label{sec: Ana}
Two mathematical analyses are provided in this section. The one is calculating the upper bound for a defined error and the other is  investigating the mean convergence of the algorithm which will be discussed later.

\subsection{Upper bound for the error}
\label{sec: Upperbound}
In this part, to ensure that the privacy-preserving DP-PGCDLMS algorithm performance is near the performance of the non-private PGCDLMS, we calculate an upper bound on the $l_2$-norm of the error vector $\Deltab\omegab=\bar{\omegab}_{k,i}-\omegab_{k,i}$, where $\bar{\omegab}_{k,i}$ is the estimator of DP-PGCDLMS and $\omegab_{k,i}$ is the estimator of PGCDLMS algorithm. So, we want to find the upper bound $\mathrm{C}_{\mathrm{max}}$ in which we have $\mathrm{D}_2=||\Deltab\omegab||^2_2=||\bar{\omegab}_{k,i}-\omegab_{k,i}||^2_2\le\mathrm{D}_{2,\mathrm{max}}$. In this regard, from (\ref{eq: decvec}), we can write
\begin{equation}
\label{eq: decvec2}
\tilde{\tilde{\phib}}_{l,i}=\tilde{\Gb}_{l,i}(\tilde{\phib}_{l,i}-\etab_{l,i}+\Vb_{l,i}).
\end{equation}
Then, substituting (\ref{eq: encvec}) into (\ref{eq: decvec2}), we have $\tilde{\tilde{\phib}}_{l,i}=\tilde{\Gb}_{l,i}\Gb^{-1}_{l,i}\phib_{l,i}+\rb_{l,i}$, where $\rb_{l,i}=\tilde{\Gb}_{l,i}\hb_{l,i}$. Then, since we have
%\begin{equation}
%\label{eq: decvec1}
%\tilde{\tilde{\phib}}_{l,i}=\tilde{\Gb}_{l,i}\Gb^{-1}_{l,i}\phib_{l,i}+\rb_{l,i},
%\end{equation}

\begin{equation}
\label{eq: omegabar}
\bar{\omegab}_{k,i}=\sum_{l\in \mathcal N_k}a_{l,k}\tilde{\tilde{\phib}}_{l,i}=\sum_{l\in \mathcal N_k}a_{l,k}\tilde{\Gb}_{l,i}\Gb^{-1}_{l,i}\phib_{l,i}+\nb_{k,i},
\end{equation}
where $\nb_{k,i}\triangleq\sum_{l\in \mathcal N_k}a_{l,k}\rb_{l,i}$, and $\omegab_{k,i}=\sum_{l\in \mathcal N_k}a_{l,k}\phib_{l,i}$, So, the error vector $\Deltab\omegab$ can be written as
%\begin{equation}
%\omegab_{k,i}=\sum_{l\in \mathcal N_k}a_{l,k}\phib_{l,i}.
%\end{equation}
\begin{equation}
\sum_{l\in \mathcal N_k}a_{l,k}(\tilde{\tilde{\phib}}_{l,i}-\phib_{l,i})=\sum_{l\in \mathcal N_k}a_{l,k}(\tilde{\Ib}_{l,i}-\Ib_L)\phib_{l,i}+\nb_{k,i},
\end{equation}
where $\Ib_L$ is the identity matrix with size $L\times L$, and $\tilde{\Ib}_{l,i}\triangleq \tilde{\Gb}_{l,i}\Gb^{-1}_{l,i}$. It is easy to write $\tilde{\Ib}_{l,i}=(\Gb_{l,i}+\Vb_{l,i})\Gb^{-1}_{l,i}=\Ib_L+\Vb_{l,i}\Gb^{-1}_{l,i}$. So, putting together, the error vector $\Deltab\omegab$ is simplified to
\begin{equation}
\label{eq: deltaomeg}
\Deltab\omegab=\sum_{l\in \mathcal N_k}a_{l,k}\Vb_{l,i}\Gb^{-1}_{l,i}\phib_{l,i}.
\end{equation}
Hence, from (\ref{eq: deltaomeg}),  $\mathrm{D}_2=||\Deltab\omegab||^2_2=(\Deltab\omegab)^T\Deltab\omegab$ can be expanded as $\mathrm{D}_2=\sum_{l\in\mathcal N_k}\sum_{l^{'}\in\mathcal N_k}a_{l,k}a_{l^{'},k}\phib^T_{l,i}\Gb^{-1}_{l,i}\Vb^T_{l,i}\Vb^T_{l^{'},i}\Gb^{-1}_{l^{'},i}\phib_{l^{'},i}$. To find an upper bound, we can use the triangle inequality. Then, we have
%\begin{equation}
%\mathrm{D}_2=\sum_{l\in\mathcal N_k}\sum_{l^{'}\in\mathcal N_k}a_{l,k}a_{l^{'},k}\phib^T_{l,i}\Gb^{-1}_{l,i}\Vb^T_{l,i}\Vb^T_{l^{'},i}\Gb^{-1}_{l^{'},i}\phib_{l^{'},i}.
%\end{equation}
\begin{equation}
\label{eq: D2}
\mathrm{D}_2=|\mathrm{D}_2|\le\sum_{l\in\mathcal N_k}\sum_{l^{'}\in\mathcal N_k}a_{l,k}a_{l^{'},k}|\phib^T_{l,i}\Gb^{-1}_{l,i}\Vb^T_{l,i}\Vb^T_{l^{'},i}\Gb^{-1}_{l^{'},i}\phib_{l^{'},i}|.
\end{equation}
If we define $\fb^T\triangleq\phib^T_{l,i}\Gb^{-1}_{l,i}$ and $\gb\triangleq\Vb^T_{l,i}\Vb^T_{l^{'},i}\Gb^{-1}_{l^{'},i}\phib_{l^{'},i}$, using the Cauchy-Schwartz inequality of vector norms i.e. $|\fb^T\gb|\le||\fb||.||\gb||$ and $||\Ab\gb||\le||\Ab||.||\gb||$, we have
\begin{equation}
\label{eq: fb}
||\fb||=||\Gb^{-1}_{l,i\phib_{l,i}}||\le||\Gb^{-1}_{l,i}||.||\phib_{l,i}||=||\Gb^{-1}_{l,i}||,
\end{equation}
and
\begin{equation}
\label{eq: gb}
||\gb||=||\Vb^T_{l,i}\Vb^T_{l^{'},i}\Gb^{-1}_{l^{'},i}\phib_{l^{'},i}||\le||\Vb^T_{l,i}||.||\Vb^T_{l^{'},i}||.||\Gb^{-1}_{l^{'},i}||,
\end{equation}
where it is assumed for simplicity that $||\phib_{l,i}||=1$, which is assured by a normalization step in the transmission step. Putting (\ref{eq: D2}), (\ref{eq: fb}), and (\ref{eq: gb}) all together and using the triangle inequality, we conclude that
\begin{displaymath}
\mathrm{D}_2\le\sum_{l\in\mathcal N_k}\sum_{l^{'}\in\mathcal N_k}a_{l,k}a_{l^{'},k}||\Gb^{-1}_{l,i}||.||\Vb^T_{l,i}||.||\Vb^T_{l^{'},i}||.||\Gb^{-1}_{l^{'},i}||=
\end{displaymath}
\begin{equation}
\Big(\sum_{l\in\mathcal N_k}a_{l,k}||\Gb^{-1}_{l,i}||.||\Vb^T_{l,i}||\Big)^2=\mathrm{D}_{2,\mathrm{max}}.
\end{equation}

\subsection{Mean convergence performance}
\label{sec: Meanconv}
In this subsection, the mean convergence of the proposed DP-PGCDLMS is investigated under some assumption which will be presented in the following. Also, a complex sufficient condition is derived. Moreover, a more simple sufficient condition on the value of step-size $\mu_k$ is derived. The assumptions which will be examined and verified experimentally are:
\begin{itemize}
\item Assumption 1: The uncorrelatedness between $\Vb_{l,i}$ and $\Gb^{-1}_{l,i}\bar{\omegab}_{l,i}$.
\item Assumption 2: $\Vb_{l,i}$ and $\Gb^{-1}_{l,i}\pb_{l,i}$ are uncorrelated, and $\mathrm{E}\{\Vb_{l,i}\}=0$.
\end{itemize}

 We define the error vector as
\begin{equation}
\label{eq: tilomega}
\tilde{\omegab}_{k,i}=\bar{\omegab}_{k,i}-\omegab^o.
\end{equation}
Neglecting the noise term $\nb_{k,i}$ of (\ref{eq: omegabar}), and using $\phib_{l,i}=\bar{\omegab}_{l,i}+\mu_l\pb_{l,i}$, by replacing (\ref{eq: omegabar}) into (\ref{eq: tilomega}), we have
\begin{equation}
\label{eq: tilomeg1}
\tilde{\omegab}_{k,i+1}=\sum_{l\in \mathcal N_k}a_{l,k}\tilde{\Gb}_{l,i}\Gb^{-1}_{l,i}(\bar{\omegab}_{l,i}+\mu_l\pb_{l,i})-\omegab^o.
\end{equation}
Now, writing $\omegab^o=\sum_{l\in\mathcal N_k}a_{l,k}\omegab^o$,  and expanding (\ref{eq: tilomeg1}), and using $\tilde{\Gb}_{l,i}\Gb^{-1}_{l,i}=\Ib_L+\Vb_{l,i}\Gb^{-1}_{l,i}$, we derive
\begin{displaymath}
\tilde{\omegab}_{k,i+1}=\sum_{l\in\mathcal N_k}a_{l,k}\tilde{\omegab}_{l,i}+\sum_{l\in\mathcal N_k}a_{l,k}\Vb_{l,i}\Gb^{-1}_{l,i}\bar{\omegab}_{l,i}
\end{displaymath}
\begin{equation}
\label{eq: finalomeg}
+\mu_k\sum_{l\in\mathcal N_k}a_{l,k}\tilde{\Gb}_{l,i}\Gb^{-1}_{l,i}\pb_{l,i}.
\end{equation}
Now, taking the expectation operator $\mathrm{E}\{.\}$ form both sides of (\ref{eq: finalomeg}), we have
\begin{displaymath}
\tilde{\tilde{\omegab}}_{k,i+1}\triangleq\mathrm{E}\{\tilde{\omegab}_{k,i+1}\}=\sum_{l\in\mathcal N_k}a_{l,k}\tilde{\tilde{\omegab}}_{l,i}
\end{displaymath}
\begin{equation}
\label{eq: ff}
+\sum_{l\in\mathcal N_k}a_{l,k}\mathrm{E}\{\Vb_{l,i}\}\mathrm{E}\{\Gb^{-1}_{l,i}\bar{\omegab}_{l,i}\}+\mu_k\sum_{l\in\mathcal N_k}a_{l,k}\mathrm{E}\{\tilde{\Gb}_{l,i}\Gb^{-1}_{l,i}\pb_{l,i}\},
\end{equation}
where assumption 1 is used. Defining the expectation of the third term of (\ref{eq: ff}) which is $\mathrm{T}_3=\mathrm{E}\{\tilde{\Gb}_{l,i}\Gb^{-1}_{l,i}\pb_{l,i}\}$, we obtain
\begin{displaymath}
\mathrm{T}_3=\mathrm{E}\{(\Ib_L+\Vb_{l,i}\Gb^{-1}_{l,i})\pb_{l,i}\}=\mathrm{E}\{\pb_{l,i}\}+\mathrm{E}\{\Vb_{l,i}\Gb^{-1}_{l,i}\pb_{l,i}\}
\end{displaymath}
\begin{equation}
\label{eq: T3}
=\mathrm{E}\{\pb_{l,i}\}+\mathrm{E}\{\Vb_{l,i}\}\mathrm{E}\{\Gb^{-1}_{l,i}\pb_{l,i}\}=\mathrm{E}\{\pb_{l,i}\},
\end{equation}
where assumption 2 is used. To calculate $\mathrm{E}\{\pb_{l,i}\}$, we write
\begin{equation}
\label{eq: pb1}
\pb_{l,i}=\sum_{l^{'}\in\mathcal N_k}a_{l^{'},l}\ub_{l^{'},i}(d_{l^{'},i}-\ub^T_{l^{'},i}\bar{\omegab}_{l^{'},i}).
\end{equation}
Replacing $d_{l^{'},i}=\ub^T_{l^{'},i}\omegab^o+\etab_{l^{'},i}$ into (\ref{eq: pb1}), we then have
\begin{equation}
\label{eq: pb2}
\pb_{l,i}=\sum_{l^{'}\in\mathcal N_k}a_{l^{'},l}\ub_{l^{'},i}\ub^T_{l^{'},i}(\omegab^o-\bar{\omegab}_{l^{'},i}).
\end{equation}
Taking the expectation of both sides of (\ref{eq: pb2}), we reach
\begin{equation}
\label{eq: Epb}
\mathrm{E}\{\pb_{l,i}\}=-\sum_{l^{'}\in\mathcal N_k}a_{l^{'},l}\Rb_{l^{'},l}\tilde{\tilde{\omegab}}_{l^{'},i},
\end{equation}
where the covariance matrix $\Rb_{l^{'},l}\triangleq \mathrm{E}\{\ub_{l^{'},i}\ub^T_{l^{'},i}\}$.
Now, substituting (\ref{eq: Epb}) and (\ref{eq: T3}) into (\ref{eq: ff}), and from assumption of being zero mean of $\Vb_{l,i}$, we find that
\begin{displaymath}
\tilde{\tilde{\omegab}}_{k,i+1}=\sum_{l\in\mathcal N_k}a_{l,k}\tilde{\tilde{\omegab}}_{l,i}-\mu_k\sum_{l\in\mathcal N_k}a_{l,k}\sum_{l^{'}\in\mathcal N_l}a_{l^{'},l}\Rb_{l^{'},l}\tilde{\tilde{\omegab}}_{l,i}
\end{displaymath}
\begin{equation}
=\sum_{l\in\mathcal N_k}a_{l,k}(\Ib_L-\mu_k\sum_{l^{'}\in\mathcal N_l}a_{l^{'},l}\Rb_{l^{'},l})\tilde{\tilde{\omegab}}_{l,i}.
\end{equation}
If we define $\Bb_{l}\triangleq \sum_{l^{'}\in\mathcal N_l}a_{l^{'},l}\Rb_{l^{'},l}$, then we have the following recursion formula for $\tilde{\tilde{\omegab}}_{k,i+1}$:
\begin{equation}
\label{eq:gb0}
\tilde{\tilde{\omegab}}_{k,i+1}=\sum_{l\in\mathcal N_k}a_{l,k}(\Ib_L-\mu_k\Bb_{l})\tilde{\tilde{\omegab}}_{l,i}.
\end{equation}
Let us define the following global quantities of $\tilde{\tilde{\omegab}}_i= \operatorname{col}\left\{\tilde{\tilde{\omegab}}_{1, i}, \ldots, \tilde{\tilde{\omegab}}_{N, i}\right\}$, $\mathcal{B}=\operatorname{diag}\left\{\mathbf{\Bb}_{1},\ldots,\mathbf{\Bb}_{N}\right\}$, $\mathcal{M}=\operatorname{diag}\left\{\mu_1 \Ib_L, \ldots, \mu_N \Ib_L\right\}$, and $\mathcal{A}=\mathbf{A} \otimes \Ib_L$,
%$$\tilde{\tilde{\omegab}}_i= \operatorname{col}\left\{\tilde{\tilde{\omegab}}_{1, i}, \ldots, \tilde{\tilde{\omegab}}_{N, i}\right\},$$
%$$\mathcal{B}=\operatorname{diag}\left\{\mathbf{\Bb}_{1},\ldots,\mathbf{\Bb}_{N}\right\},$$
%$$\mathcal{M}=\operatorname{diag}\left\{\mu_1 \Ib_L, \ldots, \mu_N \Ib_L\right\},$$
%$$\mathcal{A}=\mathbf{A} \otimes \Ib_L,$$
where $(\mathbf{A})_{l, k}=a_{l, k}$. Note that operators $\operatorname{col}\{\cdot\}$ and $\otimes$ denote the vectorization operation and the Kronecker product, respectively. Then (\ref{eq:gb0}) can be rewritten as:
\begin{equation}
\label{eq:gb1}
\tilde{\tilde{\omegab}}_{i+1}=\mathcal{A}^{\mathrm{T}}\left(\Ib_{L N}-\mathcal{M} \mathcal { B }\right) \tilde{\tilde{\omegab}}_i,
\end{equation}
%or
%\begin{equation}
%\label{eq:gb2}
%E\{\tilde{\omegab}_{i+1}\}=\mathcal{A}^{\mathrm{T}}\left(\Ib_{L N}-\mathcal{M} \mathcal { B }\right) E\{\tilde{\omegab}_i\},
%\end{equation}
%where $\tilde{\omegab}_i= \operatorname{col}\left\{\tilde{\omegab}_{1, i}, \ldots, \tilde{\omegab}_{N, i}\right\}$.

It is seen from (\ref{eq:gb1}) that the combination matrix $\mathcal{A}^{\mathrm{T}}$ appears pre-multiplying the block diagonal matrix $\left(\Ib_{L N}-\mathcal{M} \mathcal { B }\right)$. Employing the block maximum norm \cite{Sayed14} with blocks of size $L \times L$, we conclude that $\rho(\mathcal{F})\leq \rho(\Ib_{L N}-\mathcal{M} \mathcal { B })$, where $\mathcal{F} = \mathcal{A}^{\mathrm{T}}\left(\Ib_{L N}-\mathcal{M} \mathcal { B }\right)$ and $\rho(\cdot)$ represents the spectral radius of the matrix therein. Therefore, the matrix $\mathcal{F}$ becomes stable whenever the block-diagonal matrix $\left(\Ib_{L N}-\mathcal{M} \mathcal { B }\right)$ is stable. It is easily seen that this latter condition is guaranteed for step-sizes $\mu_{k}$ satisfying $0<\mu_{k}<\frac{2}{\rho(\mathbf{\Bb}_{l})}$ for $k,l=1,2 \ldots,N$, or simply $0<\mu_{k}<\frac{2}{\lambda_{max}(\mathbf{\Bb}_{l})}$. Using the definition $\Bb_{l}\triangleq \sum_{l^{'}\in\mathcal N_l}\Rb_{l^{'},l}$, the convergence condition is simplified to
\begin{equation}
\label{eq:gbnew1}
0<\mu_k<\frac{2}{\max _{l=1, \ldots, N} \lambda_{\max }(\Rb_{l^{'}, l})}.
\end{equation}
For the special case when the regression vectors are white, i.e., $\Rb_{l^{'},l}=\sigma^2_u\delta_{l^{'},l}$, we can express (\ref{eq:gbnew1}) as $0<\mu_k<\frac{2}{\sigma^2_{u}}$.

\section{Simulation Results}
\label{sec: Simulation}
In this section, the simulation results are presented. The network used in the simulation has $N=16$ agents, which is similar to that used in \cite{Zayy22CSSP}. The size of the unknown vector is $L=20$ and the elements are derived from a unit normal random variable with zero mean. The regression vector elements are also white unit normals with zero mean. The measurement noises are zero mean white Gaussian random variables with variances $\sigma^2_u=0.05$. The noisy AWGN channels between nodes are zero mean Gaussian random variables with variances $\sigma^2_v=0.05$. The combination coefficients $a_{l,k}$ and $c_{l,k}$ are selected based on uniform policy \cite{Sayed14}. For the performance metric, the MSD is used which is defined as $\mathrm{MSD}(\mathrm{dB})=20~\mathrm{log}_{10}(||\omegab-\omegab_o||_2)$. We examined the assumptions of mean convergence analysis via simulation experiment. We computed the correlation coefficient between random variables $A=\Vb_{l,i}$ and $B=\Gb^{-1}_{l,i}\bar{\omegab}_{l,i}$ which is $r(A,B)=0.183$. We computed the correlation coefficient between random variables $C=\Vb_{l,i}$ and $D=\Gb^{-1}_{l,i}\pb_{l,i}$ which is $r(C,D)=0.148$. We also computed the mean value of the error of reconstruction matrix which was $\mathrm{E}_{l,i}\{\Vb_{l,i}\}=0.057\approx 0$. However, the assumptions are not exactly validated by the assumptions, but they are satisfied to some extent. Performing simulations with different value of $\alpha$ and $\beta$ show that the best value of $\alpha$ for acquiring minimum final MSD is $\alpha=1.5$ and the proposed algorithm is not sensitive to value of $\beta$. Hence, we use $\alpha=1.5$ and $\beta=10$ in the simulations. In the first experiment, the Oracle DP-PGCDLMS, \textcolor{blue}{DP-PGCDLMS-Version1, DP-PGCDLMS-Version2}, PGCDLMS, and DLMS algorithms are compared in which the step-sizes are selected as 1, 0.1, 0.1, 0.05, 0.01, respectively. To investigate the tracking performance, we changed the value of unknown parameter vector abruptly at iteration index 1000. The result of MSD versus iteration number is depicted in Fig. 1. It is seen that the oracle DP-PGCDLMS, the non-private PGCDLMS, non-private DLMS have almost the same performance and the tacking capability of the proposed algorithm is acceptable. In the second experiment, the proposed DP-PGCDLMS \textcolor{blue}{algorithms are} compared with non-private RPRDLMS \cite{ZayyJ21}, Partial-Private DLMS (PPDLMS) \cite{Gogi22} in two cases of $M=0.8L$ and $M=L$, where $M$ is the compressed length. For the PP-DLMS, the step-size is selected as $\mu=0.01$. The results are shown in Fig. 2. \textcolor{blue}{It is observed that the proposed DP-PGCDLMS-Version2 exhibits faster convergence rate than DP-PGCDLMS-Version1. Additionally, both versions demonstrate lower final MSD and convergence rate than PPDLMS. Furthermore, the non-private RPRDLMS exhibits the lowest final MSD among all compared methods.}

%\begin{figure}[tb]
%\begin{center}
%\includegraphics[width=5.2cm]{topology.eps}
%\end{center}
%\caption{Network topology.}
%%\end{center}
%\label{fig_topo}
%\end{figure}

\begin{figure}[tb]
\begin{center}
\includegraphics[width=6.3cm]{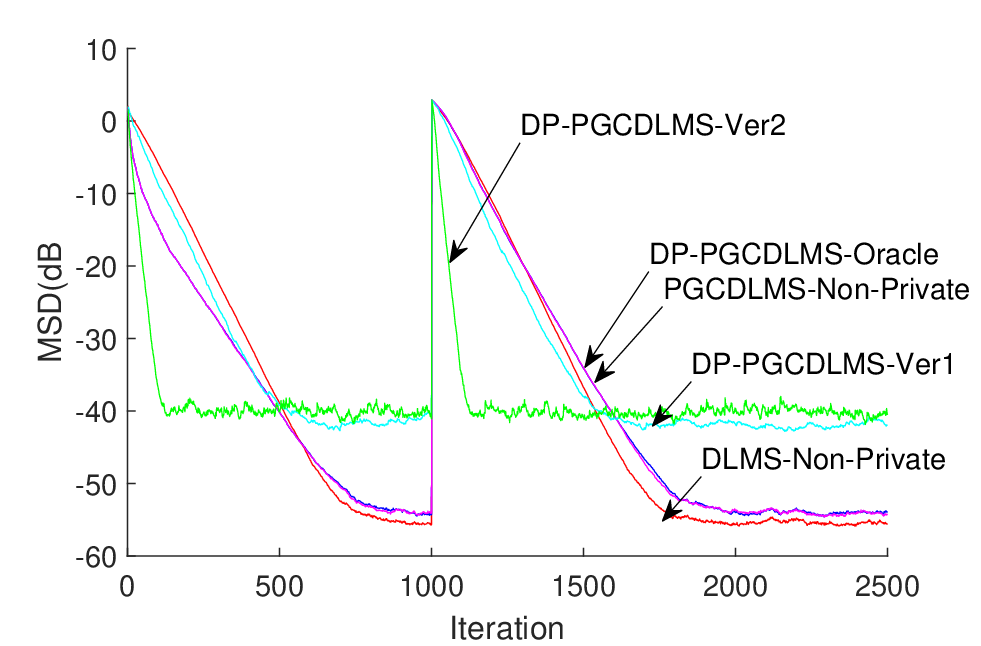}
\end{center}
\caption{MSD versus iteration number in the tracking case.}
%\end{center}
\label{fig1}
\end{figure}

\begin{figure}[tb]
\begin{center}
\includegraphics[width=6.3cm]{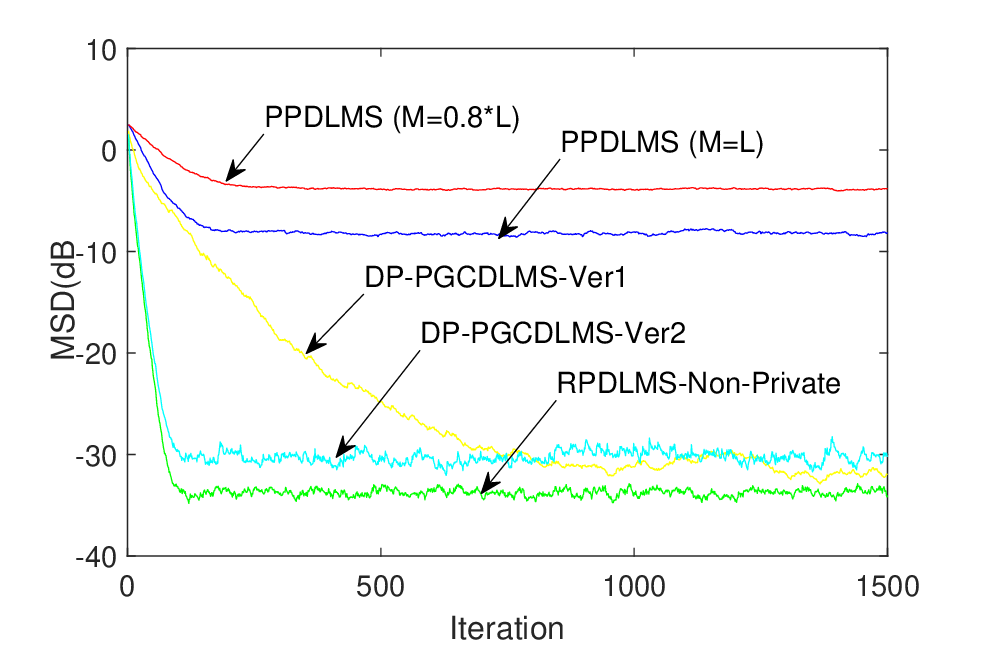}
\end{center}
\caption{MSD versus iteration number for performance comparison of various algorithms.}
%\end{center}
\label{fig2}
\end{figure}

%\begin{figure}[tb]
%\begin{center}
%\includegraphics[width=5cm]{exp1_1.eps}
%\end{center}
%\caption{MSD versus iteration number for various algorithms when $\sigma_v=0.2$.}
%%\end{center}
%\label{fig3}
%\end{figure}

\section{Conclusion and future work}
\label{sec: con}
In this paper, a privacy preserving distributed estimation algorithm is suggested which uses both cryptography-based methods and differential privacy (DP). The inverse of proportionate gain matrix in PGCDLMS is used as a key matrix to perturb the estimation to enhance the privacy. Also, DP noise is added to even yield more privacy. At the receiver of the local node, the noise is subtracted and the gain matrix is used as the key matrix to recover the intermediate estimation as a message. The benefit of using proportionate gain matrix in PGCDLMS is that it has closed form which enables us to reconstruct the key matrix without sharing the key matrix. Mathematical analysis of the proposed DP-PGCDLMS is provided in the paper. Simulation results show the effectiveness of the proposed algorithm. \textcolor{blue}{While we recognize the lack of explicit performance guarantees in our current algorithm version, we are dedicated to exploring methods to integrate these assurances in our future work.}

\end{document}